\documentclass{llncs}

\usepackage{amsmath}
\usepackage{amssymb}

\let\emptyset=\varnothing
\newcommand{\lra}{\mathop{\Longrightarrow}\limits}
\newcommand{\ie}{\emph{i.e.}}

\newcommand{\Var}{\mathit{Var}}

\newcommand{\arity}[1]{\mathit{arity}({#1})}
\newcommand{\query}[2]{\langle{#1}\,|\,{#2}\rangle}
\newcommand{\restrict}[2]{#1_{#2}}
\newcommand{\rel}[1]{\mathit{rel}({#1})}
\newcommand{\sat}[2]{\mathit{sat}({#1},{#2})}
\newcommand{\Set}[1]{\mathit{Set}({#1})}

\begin{document}

\title{An Improved Non-Termination Criterion for Binary Constraint Logic Programs}
\author{Etienne Payet \and Fred Mesnard}
\institute{IREMIA -  Universit\'e de La R\'eunion, France\\
           email: epayet@univ-reunion.fr} 
\maketitle

\begin{abstract}On  one hand, termination analysis of logic programs 
is now a fairly established research topic
within the logic programming community. On the other hand,
non-termination analysis seems to remain a much less attractive subject. 
If we divide this line of research into two kinds of approaches: dynamic
versus static analysis, this paper belongs to the latter. It proposes
a criterion for detecting non-terminating atomic queries  with respect to
binary CLP clauses, which strictly generalizes our previous works on this subject.
We give a generic operational definition and a logical form of this criterion. 
Then we show that the logical form is correct and complete with respect 
to the operational definition.
\end{abstract}


\section{Introduction}\label{section-intro}

On  one hand, termination analysis of logic programs is 
a fairly established research topic
within the logic programming community, see the surveys \cite{SD94,Mesnard03b}.
For Prolog, various tools are now available via web interfaces
and we note that the Mercury compiler, designed with industrial goals in mind
by its implementors, has included two termination analyzers (see \cite{Speirs97a}
and \cite{Fis02}) for a few years.

On the other hand, non-termination analysis seems to remain a much less attractive
subject. We can divide this line of research into two kinds of approaches:  dynamic
versus static analysis. In the former one, \cite{Bol91b} sets up some solid
foundations for loop checking, while some recent work is presented in 
\cite{Shen01a}. The main idea is to prune at runtime at least all infinite
derivations, and possibly some finite ones. In the latter approach, which
includes the work we present in this article,
\cite{DeSchreye89a,DeSchreye90a}
present an algorithm for detecting non-terminating atomic queries with
respect to a binary clause
of the type $p(\tilde{s})\leftarrow  p(\tilde{t})$. The condition is 
described  in terms of rational trees, while we aim at generalizing 
non-termination analysis for the generic CLP(X) framework.

Our analysis shares with some work on termination analysis \cite{CT99}
a key component: the binary unfoldings of a logic program \cite{Gabbrielli94a},
which transforms a finite set of definite clauses into a possibly infinite set of
facts and binary definite clauses.
While some termination analyses begin with the analysis of the 
recursive binary clauses
of an upper approximation of the binary unfoldings
of an abstract CLP(N) version of the original program, we start from a finite subset
of the binary unfoldings of the concrete program $P$ (a 
larger subset may increase the precision of the analysis, see \cite{Payet05a}
for some experimental evidence).
First we detect patterns of non-terminating atomic queries 
for binary recursive clauses and
then propagate this non-termination information to compute classes of atomic queries
for which we have a finite proof that there exists at least one infinite derivation
with respect to the subset of the binary unfoldings of $P$.

The equivalence  of termination for a program
or its binary unfoldings given in  \cite{CT99} is a corner stone of both analyses.
It allows us to conclude that any atomic query belonging to the identified above
classes of queries admits an infinite left derivation with respect to $P$.
So in this paper, we deliberately choose to restrict the analysis to binary CLP clauses
and atomic CLP queries as the result we obtain can be directly lifted to full CLP.

Our initial motivation, see \cite{Mesnard02a}, is to complement termination analysis 
with non-termination inside the logic programming paradigm 
in order to detect optimal termination conditions expressed in a language describing
classes of queries. We started from a generalization of the lifting lemma where
we may ignore some arguments. For instance, from the clause 
$p(f(X),Y) \leftarrow p(X,g(Y))$, we can conclude that the atomic query $p(X,t)$
loops for any term $t$, thus ignoring the second argument. 
Then we have extended the approach, see
\cite{Payet05a} which gives the full picture of the non-termination analysis,
an extensive experimental evaluation, and a detailed comparison with related works.
For instance, from the clause $p(f(X),g(Y)) \leftarrow p(X,g(b))$, and
with the help of the criterion designed in \cite{Payet05a}
we can now conclude
that $p(X,t)$ loops  for any term $t$ which is an instance of $g(X)$.

Although we obtained interesting experimental results from such a criterion,
the overall approach remains quite syntactic, with an \emph{ad hoc} flavor and
tight links to some basic 
logic programming machinery such as the unification algorithm. So
we moved to the constraint logic programming scheme: in \cite{Payet04b},
we started from a generic definition of the generalization of the lifting lemma
we were looking for. Such a definition was practically useless but
we were able to give a sufficient condition expressed as a logical
formula related to the constraint binary clause
$p(\tilde{x})\leftarrow c \diamond p(\tilde{y})$ under consideration.
For some constraint domains, we showed that the condition is 
also necessary. Depending on the constraint theory, 
the validity of such a condition can be automatically decided. 
Moreover, we showed that the syntactic criterion
we used in \cite{Mesnard02a} was actually equivalent to the logical criterion
and could be considered as a correct and complete
implementation specialized for the algebra
of finite trees $\mathit{Term}$. 

The main contribution of this article consists in 
a strict generalization of the logical criterion
defined in \cite{Payet04b} which allows us to reconstruct 
the syntactic approaches described in \cite{Mesnard02a} and
\cite{Payet05a}.
We emphasize the improvement with respect to \cite{Payet04b} in
Sect.~\ref{section-special-kind-filter} (see
Example~\ref{ex-stric-generalization}).

The paper 
is organized as follows.  
First, in Sect.~\ref{sect-preliminaries}, we introduce some
preliminary definitions. Then, in
Sect.~\ref{section-loop-inference-with-constraints},
we recall, using CLP terms, the subsumption test to detect looping queries.
In Sect.~\ref{section-loop-filters}, we present our generalized criterion
for detecting looping queries, whilst in Sect.~\ref{section-special-kind-filter}
we consider the connections with the results of~\cite{Payet04b}.


\section{Preliminaries}\label{sect-preliminaries}

For any non-negative integer $n$, $[1,n]$ denotes the set
$\{1,\dots,n\}$. If $n=0$, then $[1,n]=\emptyset$.
 

\subsection{First Order Formulas}

Throughout this paper, we consider a fixed, infinite and
denumerable set of variables $\mathcal{V}$.

A \emph{signature} defines a set of function and predicate
symbols and associates an \emph{arity} with each symbol.
%
If $\phi$ is a first order formula on a signature $\Sigma$
and $W:=\{X_1,\dots,X_n\}$ is
a set of variables, then $\exists_W \phi$
(resp. $\forall_W \phi$) denotes the 
formula $\exists X_1\dots \exists X_n \phi$
(resp. $\forall X_1\dots \forall X_n \phi$).
We let $\exists\phi$ (resp. $\forall\phi$) denote
the existential (resp. universal) closure of $\phi$.
%
A \emph{$\Sigma$-structure} $\mathcal{D}$ is an interpretation
of the symbols in the signature $\Sigma$.
It is a pair $(D,[\cdot])$ where $D$ is a set
called the \emph{domain} of $\mathcal{D}$ and $[\cdot]$
maps:
\begin{itemize}
\item each function symbol $f$ of arity $n$ in $\Sigma$ to
  a function $[f]: D^n \rightarrow D$,
\item each predicate symbol $p$ of arity $n$ in $\Sigma$ to
  a boolean function $[p]: D^n \rightarrow \{0,1\}$.
\end{itemize}
A \emph{$\mathcal{D}$-valuation} (or simply a \emph{valuation}
if the $\Sigma$-structure $\mathcal{D}$ is understood)
is a mapping $v:\mathcal{V}\rightarrow D$.
Every $\mathcal{D}$-valuation
$v$ extends (by morphism) to terms:
\begin{itemize}
\item $v(f(t_1,\dots,t_n)) := [f](v(t_1),\dots,v(t_n))$
  if $f(t_1,\dots,t_n)$ is a term.
\end{itemize}
A $\mathcal{D}$-valuation $v$ induces a valuation 
$[\cdot]_v$ of formulas to $\{0,1\}$:
\begin{itemize}
\item $[p(t_1,\dots,t_n)]_v := [p](v(t_1),\dots,v(t_n))$
  if $p(t_1,\dots,t_n)$ is an atomic proposition,
\item if $\phi_1$ and $\phi_2$ are formulas and
  $\circ\in\{\land,\lor,\rightarrow,\leftrightarrow\}$,
  $[\lnot\phi_1]_v$ and $[\phi_1\mathop{\circ}\phi_2]_v$ are deduced
  from $[\phi_1]_v$, $[\phi_2]_v$ and the truth table
  of $\lnot$ and $\circ$,
\item if $X$ is a variable and $\phi$ is a formula,
  $[\exists X \phi]_v=1$ if and only if there exists
  a valuation $v'$ such that $[\phi]_{v'}=1$ and
  for each variable $Y$ distinct from $X$,
  $v'(Y)=v(Y)$,
\item if $X$ is a variable and $\phi$ is a formula,
  $[\forall X \phi]_v=1$ if and only if $[\phi]_{v'}=1$
  for every valuation $v'$ such that
  for each variable $Y$ distinct from $X$,
  $v'(Y)=v(Y)$.
\end{itemize}
Given a formula $\phi$, we write
$\mathcal{D}\models_v \phi$ if $[\phi]_v=1$ and
$\mathcal{D}\not\models_v \phi$ if $[\phi]_v=0$.
We write $\mathcal{D}\models \phi$ if and only if
for every $\mathcal{D}$-valuation $v$, we have
$\mathcal{D}\models_v \phi$.
Notice that $\mathcal{D}\models \forall\phi$ if and only
if $\mathcal{D}\models \phi$,
that $\mathcal{D}\models \exists\phi$ if and only if
there exists a $\mathcal{D}$-valuation $v$ such that
$\mathcal{D}\models_v \phi$, and that
$\mathcal{D}\models \lnot\exists\phi$ if and only if
$\mathcal{D}\models \lnot\phi$.

Given a $\Sigma$-structure $\mathcal{D}$, we say that a
$\Sigma$-formula $\phi$ is \emph{satisfiable} (resp.
\emph{unsatisfiable}) in $\mathcal{D}$ if
$\mathcal{D}\models \exists\phi$ (resp.
$\mathcal{D}\models \lnot\phi$).
We say that $\mathcal{D}$ is a \emph{model}
of a set $S$ of $\Sigma$-formulas if for each element $\phi$ of
$S$ we have $\mathcal{D}\models \phi$. Given two sets $S$ and
$T$ of $\Sigma$-formulas, we say that $S$ 
\emph{semantically implies} $T$, written $S\models T$, if
every model of $S$ is also a model of $T$.

\subsection{Sequences}
Sequences of distinct variables are denoted by
$\tilde{X}$, $\tilde{Y}$, $\tilde{Z}$, $\tilde{U}$, \dots and
sequences of (not necessarily distinct) terms are
denoted by $\tilde{s}$, $\tilde{t}$, \dots
Given two sequences of $n$ terms
$\tilde{s}:=(s_1,\dots,s_n)$ and
$\tilde{t}:=(t_1,\dots,t_n)$, we write
$\tilde{s}=\tilde{t}$ either to denote the formula
$s_1=t_1 \land \dots \land s_n=t_n$ or as a
shorthand for ``$s_1=t_1$ and \dots{} and $s_n=t_n$".
Moreover, given a valuation $v$, we write
$v(\tilde{s})$ to denote the sequence
$(v(s_1),\dots,v(s_n))$. Finally, given a
sequence $\tilde{X}:=(X_1,\dots,X_n)$ of
distinct variables and given a formula $\phi$,
we write $\exists_{\tilde{X}} \phi$
(resp. $\forall_{\tilde{X}} \phi$) to denote the
formula $\exists X_1 \dots \exists X_n \phi$
(resp $\forall X_1 \dots \forall X_n \phi$).

\subsection{Constraint Domains}
We  recall  some basic  definitions about CLP, see~\cite{JMMS98}
for more details.
In this paper, we consider a  constraint  logic  programming
language CLP($\mathcal{C}$) based on the  constraint  domain
$\mathcal{C} := \langle \Sigma_{\mathcal{C}},
\mathcal{L}_{\mathcal{C}}, \mathcal{D}_{\mathcal{C}},
\mathcal{T}_{\mathcal{C}}, \mathit{solv}_{\mathcal{C}}\rangle$.
%
The constraint domain signature $\Sigma_{\mathcal{C}}$ is a pair
$\langle F_{\mathcal{C}}, \Pi_{\mathcal{C}}\rangle$
where $F_{\mathcal{C}}$ is a set of function symbols
and $\Pi_{\mathcal{C}}$ is a set of predicate symbols.
%
%
The domain of computation $\mathcal{D}_{\mathcal{C}}$
is a $\Sigma_{\mathcal{C}}$-structure
$(D_\mathcal{C},[\cdot]_{\mathcal{C}})$
that is the intended interpretation of the constraints.
We assume the following:
\begin{itemize}
\item $\mathcal{C}$ is ideal,
\item the predicate symbol $=$ is in $\Sigma_{\mathcal{C}}$ and
  is interpreted as identity in $D_{\mathcal{C}}$,
\item $\mathcal{D}_{\mathcal{C}}$ and $\mathcal{T}_{\mathcal{C}}$
  correspond on $\mathcal{L}_{\mathcal{C}}$,
\item $\mathcal{T}_{\mathcal{C}}$ is satisfaction complete
  with respect to $\mathcal{L}_{\mathcal{C}}$,
\item the theory and the solver agree in the sense that for
  every $c\in \mathcal{L}_{\mathcal{C}}$,
  $\mathit{solv}_{\mathcal{C}}(c) = \mathtt{true}$
  if and only if
  $\mathcal{T}_{\mathcal{C}} \models \exists c$.
  Consequently, as $\mathcal{D}_{\mathcal{C}}$ and
  $\mathcal{T}_{\mathcal{C}}$ correspond on
  $\mathcal{L}_{\mathcal{C}}$, we have, for every
  $c\in \mathcal{L}_{\mathcal{C}}$,
  $\mathit{solv}_{\mathcal{C}}(c) = \mathtt{true}$
  if and only if
  $\mathcal{D}_{\mathcal{C}} \models \exists c$.
\end{itemize}
\begin{example}[$\mathcal{R}_{\mathit{lin}}$]
  \label{example-reals}
  The constraint domain $\mathcal{R}_{\mathit{lin}}$
  has $<$, $\leq$, $=$, $\geq$ and $>$ as predicate symbols,
  $+$, $-$, $*$, $/$  as function symbols and sequences of
  digits (possibly with a decimal point) as constant symbols.
  Only linear constraints are admitted. The domain of
  computation is the structure with reals as domain and
  where the predicate symbols and the function symbols are
  interpreted as the usual relations and functions over reals.
  The theory $\mathcal{T}_{\mathcal{R}_{\mathit{lin}}}$ is the
  theory of real closed fields~\cite{Sho67}.
  A constraint solver for $\mathcal{R}_{\mathit{lin}}$ always
  returning either {\tt true} or {\tt false} is described
  in~\cite{RVH96}.
\end{example}

\begin{example}[Logic Programming]
  \label{example-lp}
  The constraint domain $\mathit{Term}$ has $=$ as
  predicate symbol and strings of alphanumeric
  characters as function symbols.
  The domain of computation of $\mathit{Term}$ is
  the set of \emph{finite trees} (or, equivalently,
  of finite terms), $\mathit{Tree}$, while the theory
  $\mathcal{T}_{\mathit{Term}}$ is Clark's equality
  theory~\cite{Cla78}.
  The interpretation of a constant is a tree with a
  single node labeled with the constant. The
  interpretation of an $n$-ary function symbol $f$
  is the function
  $f_{\mathit{Tree}}:\mathit{Tree}^n \rightarrow
  \mathit{Tree}$ mapping the trees $T_1$, \dots, $T_n$
  to a new tree with root labeled with $f$ and with
  $T_1$, \dots, $T_n$ as child nodes. A constraint
  solver always returning either {\tt true} or {\tt false}
  is provided by the \emph{unification} algorithm.
  CLP($\mathit{Term})$ coincides then with logic programming.
\end{example}

\subsection{Operational Semantics}
The signature in which all programs and queries
under consideration are included is
$\Sigma_L := \langle F_L, \Pi_L\rangle$ with
$F_L := F_{\mathcal{C}}$ and
$\Pi_L := \Pi_{\mathcal{C}} \cup \Pi'_L$ where
$\Pi'_L$, the set of predicate symbols that can
be defined in programs, is disjoint from
$\Pi_{\mathcal{C}}$.
We assume that each predicate symbol $p$ in
$\Pi_L$ has a unique arity denoted by
$\arity{p}$.

An \emph{atom} has the form $p(\tilde{t})$ where
$p\in\Pi'_L$  and $\tilde{t}$
is a sequence of $\arity{p}$ $\Sigma_L$-terms.
Throughout this paper, when we write $p(\tilde{t})$,
we implicitly assume that $\tilde{t}$ contains
$\arity{p}$ terms.
%
A CLP($\mathcal{C}$) \emph{program} is a finite set
of rules. A \emph{rule} has the form
$H \leftarrow c \diamond B$
where $H$ and $B$ are atoms and
$c$ is a finite conjunction of primitive
constraints such that $\mathcal{D}_{\mathcal{C}}
\models \exists c$.
A \emph{query} has the form $\query{A}{d}$
where $A$ is an atom and $d$ is a finite conjunction
of primitive constraints.
Given an atom $A:=p(\tilde{t})$, we write
$\rel{A}$ to denote the predicate symbol $p$.
Given a query $S:=\query{A}{d}$,
we write $\rel{S}$ to denote the predicate
symbol $\rel{A}$.
The set of variables occurring in
some syntactic objects $O_1,\dots,O_n$ is denoted
$\Var(O_1,\dots,O_n)$.


The examples of this paper make use of
the language CLP($\mathcal{R}_{\mathit{lin}}$)
and the language CLP($\mathit{Term}$).
In program and query examples,
variables begin with an upper-case
letter, $[\mathit{Head}|\mathit{Tail}]$ denotes a list with
head $\mathit{Head}$ and tail $\mathit{Tail}$, and
$[\,]$ denotes an empty list.

We consider the following operational semantics given in
terms of \emph{derivations} from queries to queries.
Let $\query{p(\tilde{u})}{d}$ be a query
and $r:= p(\tilde{s})\leftarrow c \diamond q(\tilde{t})$ be a rule.
Let $r' := p(\tilde{s}')\leftarrow c' \diamond q(\tilde{t}')$
be a variant of $r$ variable disjoint with
$\query{p(\tilde{u})}{d}$ such that
$\mathit{solv}_{\mathcal{C}}(\tilde{s}' = \tilde{u}
\land c' \land d) = \mathtt{true}$. Then,
$\query{p(\tilde{u})}{d} \lra_r
\query{q(\tilde{t}')}{\tilde{s}'=\tilde{u}\land c'\land d}$
is a  \emph{derivation step} of $\query{p(\tilde{u})}{d}$
with respect to $r$ with $r'$ as its
\emph{input rule}.
We write $S \lra_P^+ S'$ to summarize a finite number
($> 0$) of derivation steps from $S$ to $S'$
where each input rule is a variant of a rule from
program $P$. 
Let $S_0$ be a query. A sequence of derivation steps
$S_0 \lra_{r_1} S_1 \lra_{r_2} \cdots$ of maximal
length is called a \emph{derivation}
of $P\cup\{S_0\}$ if $r_1$, $r_2$, \dots are rules from $P$
and if the \emph{standardization apart} condition holds, \ie{}
each input rule used is variable disjoint from the
initial query $S_0$ and from the input rules
used at earlier steps.
We say $S_0$ \emph{loops} with respect to $P$
if there exists an infinite derivation of $P\cup\{S_0\}$.


\section{Loop Inference with Constraints}
\label{section-loop-inference-with-constraints}
In the logic programming framework, the subsumption test provides
a simple way to infer looping queries: if, in a logic
program $P$, there is a rule $p(\tilde{s})\leftarrow p(\tilde{t})$
such that $p(\tilde{t})$ is more general than $p(\tilde{s})$,
then the query $p(\tilde{s})$ loops with respect to $P$.
In this section, we extend this result to the constraint logic
programming framework.

\subsection{A ``More General Than" Relation}
\label{section-set-query}

A query can be viewed as a finite description of a possibly
infinite set of atoms, the arguments of which are values
from $D_{\mathcal{C}}$. 
%
\begin{example}
  Suppose that $\mathcal{C} = \mathcal{R}_{\mathit{lin}}$.
  \begin{itemize}
  \item The query $\query{p(2*X)}{X\geq -1}$ describes
    those atoms $p(x)$ where $x$ is a real and the term $2*X$
    can be made equal to $x$ while the constraint $X\geq -1$
    is satisfied.
  \item The query $\query{q(X,Y)}{Y\leq X+2}$ describes
    those atoms $q(x,y)$ where $x$ and $y$ are reals and
    $X$ and $Y$ can be made equal to $x$ and $y$ respectively
    while the constraint $Y\leq X+2$ is satisfied.
  \end{itemize}
\end{example}
In order to capture this intuition, we introduce the
following definition.
%
\begin{definition}[Set Described by a Query]
  The set of atoms that is described by a query
  $S:=\query{p(\tilde{t})}{d}$ is denoted by
  $\Set{S}$ and is defined as:
  $\Set{S} = \{p(v(\tilde{t})) \; | \; \mathcal{D}_{\mathcal{C}}
  \models_v d\}$.
\end{definition}

Clearly, $\Set{\query{p(\tilde{t})}{d}}=\emptyset$ if and
only if $d$ is unsatisfiable in $\mathcal{D}_{\mathcal{C}}$.
Moreover, two variants describe the same set.
Notice that the operational semantics we introduced above
can be expressed using sets described by queries:
\begin{lemma}
  \label{lemma-operational-sem}
  Let $S$ be a query and
  $r:=H\leftarrow c\diamond B$ be a rule.
  There exists a derivation step of $S$ with respect to $r$
  if and only if  $\Set{S}\cap\Set{\query{H}{c}}\neq\emptyset$.
\end{lemma}

The ``more general than" relation we consider is defined as follows:
%
\begin{definition}[More General]
  We say that a query $S'$ is \emph{more general than}
  a query $S$ if $\Set{S}\subseteq \Set{S'}$.
\end{definition}
%
\begin{example}\hspace{0cm}
  \begin{itemize}
  \item In any constraint domain, $\query{p(X)}{\mathit{true}}$
    is more general than any query $S$ verifying
    $\mathit{rel}(S)=p$;
  \item In the constraint domain $\mathit{Term}$, the query
    $\query{p(Y)}{Y=f(X)}$ is more general than the query
    $\query{p(Y)}{Y=f(f(X))}$;
  \item In the constraint domain $\mathcal{R}_{\mathit{lin}}$,
    the query $\query{q(X,Y)}{Y\leq X+2}$
    is more general than the query
    $\query{q(X,Y)}{Y\leq X+1}$.
  \end{itemize}
\end{example}

\subsection{Loop Inference}
\label{section-loop-constraints}

Suppose we have a derivation step $S\lra_r T$ where
$r:=H\leftarrow c\diamond B$. Then,
by Lemma~\ref{lemma-operational-sem},
$\Set{S}\cap\Set{\query{H}{c}}\neq\emptyset$.
Hence, if $S'$ is a query that is more general than
$S$, as $\Set{S}\subseteq\Set{S'}$, we have
$\Set{S'}\cap\Set{\query{H}{c}}\neq\emptyset$. So,
by Lemma~\ref{lemma-operational-sem}, there exists
a query $T'$ such that $S'\lra_r T'$.
The following lifting result says that, moreover, $T'$ is
more general than $T$: 
\begin{theorem}[Lifting]\label{theorem-lifting}
  Consider a derivation step $S \lra_r T$ and
  a query $S'$ that is more general than $S$.
  Then, there exists a derivation step $S' \lra_r T'$
  where $T'$ is more general than $T$.
\end{theorem}

From this theorem, we derive two corollaries that
can be used to infer looping queries just from the
text of a CLP($\mathcal{C}$) program:

\begin{corollary}
  \label{coro-p-if-p}
  Let $r:=H\leftarrow c\diamond B$ be a rule.
  If $\query{B}{c}$ is more general
  than $\query{H}{c}$ then
  $\query{H}{c}$ loops with respect to $\{r\}$.
\end{corollary}
%
\begin{corollary}
  \label{coro-p-if-q}
  Let $r:=H\leftarrow c\diamond B$
  be a rule from a program $P$.
  If $\query{B}{c}$ loops with respect to
  $P$ then $\query{H}{c}$ loops with respect to $P$.
\end{corollary}

\begin{example}\label{ex-loop-inference-append}
  Consider the CLP($\mathit{Term}$) rule $r$:
  \[\mathit{append}([X|\mathit{Xs}],\mathit{Ys},
  [X|\mathit{Zs}]) \leftarrow \mathit{true} \diamond
  \mathit{append}(\mathit{Xs},\mathit{Ys},\mathit{Zs})\]
  We note that the query 
  $\query{\mathit{append}(\mathit{Xs},\mathit{Ys},\mathit{Zs})}{\mathit{true}}$  
  is more general than the query
  $S:=\query{\mathit{append}([X|\mathit{Xs}],\mathit{Ys},
  [X|\mathit{Zs}])}{\mathit{true}}$.
  So, by Corollary~\ref{coro-p-if-p}, $S$
   loops with respect to
  $\{r\}$. Therefore, there exists an infinite derivation $\xi$ of
  $\{r\}\cup\{S\}$.
  Then, if $S'$ is a query that is more general than
  $S$, by successively
  applying the Lifting Theorem~\ref{theorem-lifting}
  to each step of $\xi$, one
  can construct an infinite derivation of
  $\{r\}\cup\{S'\}$. So, $S'$ also loops with respect to
  $\{r\}$.
\end{example}
%


\section{Loop Inference Using Filters}
\label{section-loop-filters}

The condition provided by
Corollary~\ref{coro-p-if-p} is rather weak because it fails
at inferring looping queries in some simple cases.
This is illustrated by the following example.

\begin{example}
  \label{example-neutral-Rlin}
  Assume $\mathcal{C}=\mathcal{R}_{\mathit{lin}}$.
  Let
  \[r :=p(X,Y) \leftarrow X\geq 0 \land Y\leq 10\diamond p(X+1,Y+1)\;.\]
  We have the infinite derivation:
  \[\begin{array}{rcl}
    \query{p(X,Y)}{c} & \lra_r &
    \query{p(X_1+1,Y_1+1)}{c \land c_1}\\
    & \lra_r & \query{p(X_2+1,Y_2+1)}{c \land c_1 \,\land c_2}\\
    & \vdots &
  \end{array}\]
  where:
  \[\begin{array}{lll}
     c & \text{ is the constraint } & X\geq 0\land Y\leq 10,\\
     c_1 & \text{ is the constraint } & X_1=X \land Y_1=Y \land
     X_1\geq 0 \land Y_1\leq 10 \text{ and}\\
     c_2 & \text{ is the constraint } & X_2=X_1+1 \land Y_2=Y_1+1 \land X_2\geq 0
     \land Y_2\leq 10.
  \end{array}\]
  But as in $r$, $\query{p(X+1,Y+1)}{c}$ is not more
  general than $\query{p(X,Y)}{c}$,
  Corollary~\ref{coro-p-if-p} does not allow to
  infer that $\query{p(X,Y)}{c}$ loops with respect to $\{r\}$.
\end{example}

In this section, we extend the relation ``is more general''.
Instead of comparing atoms in all positions using the ``more general''
relation, we distinguish some predicate argument positions
for which we just require that a certain property must hold, while
for the other positions we use the ``more general'' relation as
before.
Doing so, we aim at inferring more looping queries.

\begin{example}[Example \ref{example-neutral-Rlin} continued]
  \label{example-neutral-Rlin-continued}
  Let us consider argument position 1 of predicate symbol $p$.
  In the rule $r$, the argument of $p(X,Y)$ in position 1 is
  $X$ and the argument of $p(X+1,Y+1)$ in position 1 is $X+1$.
  Notice that the condition on $X$ in $c$ is $X\geq 0$
  and that if $X\geq 0$ then $X+1\geq 0$. Hence, let us define
  the condition $\delta$ as: a query satisfies
  $\delta$ if it has the form $\query{p(t_1,t_2)}{d}$ where $t_1$
  and $t_2$ are some terms and
  $\{v(t_1) \;|\; \mathcal{D}_{\mathcal{C}} \models_v d\}$ 
  is included in the set of positive real numbers.
  Then, both $S:=\query{p(X,Y)}{c}$ and
  $T:=\query{p(X+1,Y+1)}{c}$ satisfy $\delta$.
  
  So, if we consider a ``more general than" relation where we ``filter"
  queries  using $\delta$, as $S$ and $T$ both satisfy $\delta$
  and as the ``piece" $\query{p(Y+1)}{c}$ of $T$ is more general
  than the ``piece" $\query{p(Y)}{c}$ of $S$,
  by an extended version of Corollary~\ref{coro-p-if-p} we could
  infer that $S$ loops with respect to $\{r\}$.
\end{example}

\subsection{Sets of Positions}
\label{section-sets-of-pos}
A basic idea in Example~\ref{example-neutral-Rlin-continued}
lies in identifying argument positions of predicate symbols.
Below, we introduce a formalism to do so.

\begin{definition}[Set of Positions]
  A \emph{set of positions}, denoted by $\tau$,
  is a function that maps each predicate
  symbol $p \in \Pi'_L$ to a subset of
  $[1,\arity{p}]$.
\end{definition}

\begin{example}
  \label{example-set-of-pos}
  If we want to distinguish the first argument position of
  the predicate symbol $p$ defined in
  Example~\ref{example-neutral-Rlin}, we set
  $\tau := \langle p \mapsto\{1\}\rangle$.
\end{example}

\begin{definition}
  Let $\tau$ be a set of positions. Then, $\overline{\tau}$
  is the set of positions defined as: for each predicate
  symbol $p \in \Pi'_L$,
  $\overline{\tau}(p) = [1,\arity{p}]\setminus\tau(p)$.
\end{definition}

\begin{example}[Example~\ref{example-set-of-pos} continued]
  We have $\overline{\tau} = \langle p\mapsto\{2\}
  \rangle$.
\end{example}

Using a set of positions $\tau$, one can \emph{project}
syntactic objects: 
\begin{definition}[Projection] \label{def-projection}
  Let $\tau$ be a set of positions.
  \begin{itemize}
  \item Let $p\in \Pi'_L$ be a predicate symbol. The
    \emph{projection of $p$ on $\tau$} is the
    predicate symbol denoted by $\restrict{p}{\tau}$.
    Its arity equals the number of elements of $\tau(p)$.
  \item Let $p\in \Pi'_L$ be a predicate symbol of arity $n$
    and $\tilde{t}:=(t_1,\dots,t_n)$ be a sequence of
    $n$ terms. The \emph{projection of $\tilde{t}$ on
    $\tau$}, denoted by $\restrict{\tilde{t}}{\tau}$  is the
    sequence $(t_{i_1},\dots,t_{i_m})$ where
    $\{i_1,\dots,i_m\}=\tau(p)$ and
    $i_1\leq\dots\leq i_m$.
  \item Let $A:=p(\tilde{t})$ be an atom.
    The \emph{projection of $A$ on $\tau$},
    denoted by $\restrict{A}{\tau}$, is the
    atom $\restrict{p}{\tau}(\restrict{\tilde{t}}{\tau})$.
  \item The \emph{projection of a query $\query{A}{d}$ on $\tau$},
    denoted by $\restrict{\query{A}{d}}{\tau}$,
    is the query $\query{\restrict{A}{\tau}}{d}$.
  \end{itemize}
\end{definition}

\begin{example}[Example~\ref{example-neutral-Rlin}
  and Example~\ref{example-set-of-pos} continued]
  \label{example-restriction}
  The projection of the query $\query{p(X,Y)}{c}$ on $\tau$ is the query
  $\query{p_{\tau}(X)}{c}$.
\end{example}

\subsection{Filters}
According to the intuitions described in
Example~\ref{example-neutral-Rlin-continued} above, we define a filter as follows.
\begin{definition}[Filter]\label{def-filter}
  A \emph{filter}, denoted by $\Delta$, is a pair $(\tau,\delta)$
  where $\tau$ is a set of positions and $\delta$ is a
  function that maps each predicate
  symbol $p \in \Pi'_L$ to a query of the form
  $\query{\restrict{p}{\tau}(\tilde{u})}{d}$ where
  $\mathcal{D}_{\mathcal{C}} \models \exists d$
  and $\tilde{u}$ is a sequence of $\arity{\restrict{p}{\tau}}$ terms.
\end{definition}

\begin{example}[Example~\ref{example-neutral-Rlin} and
  Example~\ref{example-neutral-Rlin-continued} continued]
  \label{ex-filter-Rlin}
  Let $\delta$ be the function defined as
  $\delta := \langle \; p \mapsto \query{\restrict{p}{\tau}(X)}
  {X\geq 0} \; \rangle$. Then,
  $\Delta:=(\tau,\delta)$ is a filter.
\end{example}

\begin{example}
  \label{ex-filter-term}
  Suppose that $\mathcal{C}=\mathit{Term}$. Let
  $p \in\Pi'_L$ be a predicate symbol whose arity is 1. 
  Let $\tau:= \langle p \mapsto \{1\}\rangle$ and
  $\delta := \langle \; p \mapsto
  \query{\restrict{p}{\tau}(f(X))}
  {\mathit{true}} \; \rangle$. Then,
  $\Delta:=(\tau,\delta)$ is a filter.
\end{example}

The function $\delta$ is used to ``filter" queries as indicated
by the next definition.
\begin{definition}[Satisfies]
  \label{def-satisfies}
  Let $\Delta:=(\tau,\delta)$ be a filter and
  $S$ be a query. Let $p:=\mathit{rel}(S)$.
  We say that $S$ \emph{satisfies} $\Delta$ if
  $\Set{\restrict{S}{\tau}}\subseteq\Set{\delta(p)}$.
\end{definition}

Now we come to the extension of the relation
``more general than". 
Intuitively, $\query{p(\tilde{t'})}{d'}$ is
$\Delta$-more general than $\query{p(\tilde{t})}{d}$ if the ``more
general than" relation holds for the elements of $\tilde{t}$
and $\tilde{t'}$ whose position is not in $\tau$ while
the elements of $\tilde{t'}$ whose position is in $\tau$ satisfy
$\delta$. More formally:
\begin{definition}[$\Delta$-More General]
  \label{def-Delta-more-gen-state}
  Let $\Delta:=(\tau,\delta)$ be a filter and
  $S$ and $S'$ be two queries.
  We say that $S'$ is \emph{$\Delta$-more general than} $S$ if
  $\restrict{S'}{\overline{\tau}}$ is more general than
  $\restrict{S}{\overline{\tau}}$ and $S'$ satisfies
  $\Delta$.
\end{definition}

\begin{example}\hspace{0cm}
  \begin{itemize}
  \item In the context of Example~\ref{ex-filter-Rlin},
    $\query{p(X+1,Y+1)}{X\geq 0 \land Y\leq 10}$ is
    $\Delta$-more general than
    $\query{p(X,Y)}{X\geq 0 \land Y\leq 10}$.
  \item In the context of Example~\ref{ex-filter-term},
    $\query{p(f(f(X)))}{\mathit{true}}$ is
    $\Delta$-more general than 
    $\query{p(f(X))}{\mathit{true}}$.
  \end{itemize}
\end{example}

Notice that for any filter $\Delta:=(\tau,\delta)$ and any query $S$,
we have that $\restrict{S}{\overline{\tau}}$ is more general than itself
(because the ``more general than" relation is reflexive), but $S$ may
not satisfy $\Delta$. Hence, the ``$\Delta$-more general than" relation
is not always reflexive.
\begin{example}[Example~\ref{ex-filter-term} continued]
  $S:=\query{p(g(X))}{\mathit{true}}$ is not $\Delta$-more general
  than itself because, as
  $\Set{\restrict{S}{\tau}}=\{\restrict{p}{\tau}(g(t)) \;|\; t \text{ is a term}\}$ and
  $\Set{\delta(p)}=\{\restrict{p}{\tau}(f(t)) \;|\; t \text{ is a term}\}$,
  we have $\Set{\restrict{S}{\tau}} \cap \Set{\delta(p)} = \emptyset$.
  Hence, $S$ does not satisfy $\Delta$.
\end{example}

The fact that reflexivity does not always hold is
an expected property. Indeed,
suppose that a filter $\Delta:=(\tau,\delta)$ induces
a ``$\Delta$-more general than" relation  that is
reflexive. Then for any queries $S$ and $S'$,
we have that $S'$ is $\Delta$-more general
than $S$ if and only if $\restrict{S'}{\overline{\tau}}$ is
more general than $\restrict{S}{\overline{\tau}}$
(because, as $S'$ is $\Delta$-more general than itself,
$S'$ necessarily satisfies $\Delta$). Hence, $\delta$ is useless in the
sense that it ``does not filter anything".
Filters equipped with such a $\delta$ are studied in
Sect.~\ref{section-special-kind-filter}
and were introduced in~\cite{Payet04b} where for any
predicate symbol $p$, $\delta(p)$ is
$\query{\restrict{p}{\tau}(\tilde{X})}{\mathit{true}}$.
In this paper, we aim at generalizing the approach
of~\cite{Payet04b}. Hence, we also consider functions
$\delta$ that really filter queries.

\subsection{Derivation Neutral Filters: an Operational Definition}

In the sequel of this paper, we focus on ``derivation neutral"
filters. The name ``derivation neutral" stems from the fact
that if, in a derivation of a query $S$, we replace
$S$ by $S'$ that satisfies the filter, then we get a ``similar" derivation.
\begin{definition}[Derivation Neutral]
  \label{def-DN-filter}
  Let $r$ be a rule and $\Delta$ be a filter.
  We say that $\Delta$ is \emph{DN}
  for $r$ if for each derivation step $S \lra_r T$ and
  each query $S'$ that is $\Delta$-more general than $S$,
  there exists a derivation step $S'\lra_r T'$ 
  where $T'$ is $\Delta$-more general than $T$.
  This definition is extended to programs:
  $\Delta$ is \emph{DN} for $P$
  if it is DN for each rule of $P$.
\end{definition}

Derivation neutral filters lead to the following extended version
of Corollary~\ref{coro-p-if-p} (to get Corollary~\ref{coro-p-if-p},
take $\Delta:=(\tau,\delta)$ such that $\tau(p)=\emptyset$ for any $p$).
\begin{proposition}
  \label{propo-p-if-p-Delta}
  Let $r := H \leftarrow c \diamond B$ be a rule. 
  Let $\Delta$ be a filter that is DN for $r$.
  If $\query{B}{c}$ is $\Delta$-more
  general than $\query{H}{c}$ then
  $\query{H}{c}$ loops with respect to $\{r\}$.
\end{proposition}
\begin{example}[Example~\ref{example-neutral-Rlin-continued} continued]
  Suppose that $\Delta$ is DN for $r$.
  Now we can deduce that the query
  $\query{p(X,Y)}{X\geq 0 \land Y\leq 10}$
  loops with respect to $r$ because the query
  $\query{p(X+1,Y+1)}{X\geq 0 \land Y\leq 10}$ is
  $\Delta$-more general than the query
  $\query{p(X,Y)}{X\geq 0 \land Y\leq 10}$.
\end{example}

Computing a neutral filter from the text of a program
is not that easy if we use the definition above. The next
subsections present a logical and a syntactic characterization
that can be used to compute a filter that is DN for a given
program.

\subsection{A Logical Characterization of Derivation Neutral Filters}
\label{section-logical-characterization}
From now on, we suppose that, without loss
of generality, a rule has the form 
$p(\tilde{X}) \leftarrow c \diamond q(\tilde{Y})$ where
$\tilde{X}$ and $\tilde{Y}$ are disjoint sequences of
distinct variables. Hence, $c$ is the conjunction of
all the constraints, including unifications.
We distinguish the following set of variables
that appear inside such a rule.
\begin{definition}[Local Variables]
  Let $r:=p(\tilde{X}) \leftarrow c \diamond q(\tilde{Y})$
  be a rule. The set of \emph{local variables} of $r$
  is denoted by $\mathit{local\_var}(r)$ and is defined as:
  $\mathit{local\_var}(r) := \Var(c) \setminus (
  \Var(\tilde{X}) \cup \Var(\tilde{Y}))$.
\end{definition}

In this section, we aim at characterizing DN filters in a logical
way. To this end, we define:
\begin{definition}[sat]
  Let $S:=\query{p(\tilde{u})}{d}$ be a query and
  $\tilde{s}$ be a sequence of $\arity{p}$ terms.
  Then, $\sat{\tilde{s}}{S}$ denotes a  formula of the form
  $\exists_{\Var(S')} (\tilde{s}=\tilde{u}'\land d')$
  where $S':=\query{p(\tilde{u}')}{d'}$ is any variant of $S$
  variable disjoint with $\tilde{s}$.
\end{definition}
Clearly, the satisfiability of $\sat{\tilde{s}}{S}$ does not depend
on the choice of the variant of $S$.
Now we give a logical definition of derivation neutrality.
As we will see below, under certain circumstances, this
definition is equivalent to the operational one we
gave above.
\begin{definition}[Logical Derivation Neutral]
  \label{def-log-DN}
  We say that a filter $\Delta:=(\tau,\delta)$ is
  \emph{DNlog} for a rule
  $r:=p(\tilde{X}) \leftarrow c \diamond q(\tilde{Y})$ if
  \[\mathcal{D}_{\mathcal{C}} \models
  c \rightarrow \forall_{\restrict{\tilde{X}}{\tau}} \big[
  \sat{\restrict{\tilde{X}}{\tau}}{\delta(p)} \rightarrow
  \exists_{\mathcal{Y}}[\sat{\restrict{\tilde{Y}}{\tau}}{\delta(q)}\land c]
  \big]\]
  where $\mathcal{Y}:=\Var(\restrict{\tilde{Y}}{\tau}) \cup
  \mathit{local\_var}(r)$.
\end{definition}
Intuitively, the formula in Definition~\ref{def-log-DN} has the
following meaning. If one holds a solution $v$ for constraint $c$,
then, changing the value given to the variables of $\tilde{X}$
distinguished by $\tau$ to some value satisfying $\delta(p)$,
there exists a value for the local variables and the
variables of $\tilde{Y}$ distinguished by $\tau$ such that
$c$ is still satisfied.

%
\begin{example}
  Suppose that $\mathcal{C}=\mathcal{R}_{\mathit{lin}}$.
  Consider the rule $r:=p(X_1,X_2) \leftarrow c \diamond p(Y_1,Y_2)$
  where $c$ is the constraint
  $X_1=A+B \land A\geq 0 \land B\geq 0
    \land X_2\leq 10 \land Y_1=X_1+1\land Y_2=X_2+1$.
  Then, the local variables of $r$ are $A$ and $B$.
  Any filter $\Delta:=(\tau,\delta)$ where $\tau(p)=\{1\}$
  and $\delta(p)=\query{\restrict{p}{\tau}(X)}{X\geq 0}$ is DNlog
  for $r$. Indeed, $\restrict{\tilde{X}}{\tau}=X_1$,
  $\mathcal{Y}=\{Y_1,A,B\}$ and $\sat{t}{\delta(p)}$ is true if and
  only if $t\geq 0$.
  So the formula of Definition~\ref{def-log-DN} turns into
  $\mathcal{D}_{\mathcal{C}} \models
  c \rightarrow \forall X_1\big[X_1\geq 0 \rightarrow
  \exists_{\{Y_1,A,B\}} [Y_1\geq 0\land c]\big]$,
  which is true.
\end{example}
%
\begin{example}\label{example-DNlog-Term}
  Suppose that $\mathcal{C}=\mathit{Term}$.
  Consider the rule $r:=p(X) \leftarrow c \diamond p(Y)$
  where $c$ is the constraint
  $X=f(A)\land Y=f(f(A))$.
  Then, the only local variable of $r$ is $A$.
  Any filter $\Delta:=(\tau,\delta)$ where $\tau(p)=\{1\}$
  and $\delta(p)=\query{\restrict{p}{\tau}(X)}{X=f(A)}$ is DNlog
  for $r$. Indeed, $\restrict{\tilde{X}}{\tau}=X$,
  $\mathcal{Y}=\{Y,A\}$ and $\sat{t}{\delta(p)}$ is true if and
  only if $t$ has the form $f(\cdots)$.
  So the formula of Definition~\ref{def-log-DN} turns into
  \[\begin{array}{ll}
    \mathcal{D}_{\mathcal{C}} \models c \rightarrow \forall X
    \big[ & X \text{ has the form } f(\cdots) \\
    & \rightarrow \exists_{\{Y,A\}} [Y\text{ has the form } f(\cdots) \land c]\,\big],
    \end{array}\]
  which is true.
\end{example}

The logical definition of derivation neutrality implies the
operational one:
\begin{proposition}
  \label{prop-DNlog-implies-DN}
  Let $r$ be a rule and $\Delta$ be a filter.
  If $\Delta$ is DNlog for $r$ then
  $\Delta$ is DN for $r$.
\end{proposition}

The reverse implication does not always hold.
But when considering a special case
of  the ($SC_1$) condition of \emph{solution compactness}
given in \cite{JL87}, we get:
\begin{theorem}\label{prop-DN-implies-DNlog2}
  Let $r$ be a rule and $\Delta$ be a filter.
  Assume $\mathcal{C}$ enjoys the following property:
  for each $\alpha\in D_{\mathcal{C}}$, there exists a
  ground $\Sigma_{\mathcal{C}}$-term $a$ such that
  $[a]=\alpha$.
  Then, $\Delta$ is DN for $r$ if and only if $\Delta$ is
  DNlog for $r$.
\end{theorem}
\begin{proof}[Sketch]
  We show how the ($SC_1$) condition is used to get this result.
 
  By Proposition~\ref{prop-DNlog-implies-DN},
  we just have to establish that DN $\Rightarrow$ DNlog.
  Let $(\tau,\delta):=\Delta$ and
  $p(\tilde{X})\leftarrow c\diamond q(\tilde{Y}):=r$.
  Suppose that $\Delta$ is DN for $r$. We have to prove
  that then, the formula of Definition~\ref{def-log-DN}
  holds. Assume that $v$ is a valuation such that
  \begin{equation}\label{eq1-theo-DN-iff-DNlog}
    \mathcal{D}_{\mathcal{C}} \models_v c\;.
  \end{equation}
  By property of $\mathcal{C}$, we can consider the query
  $S:=\query{p(\tilde{a})}{\mathit{true}}$ where
  $\tilde{a}$ is a sequence of ground terms such that
  $[\tilde{a}]=v(\tilde{X})$. As $r$ and $S$ are variable
  disjoint, we have $S\lra_r T$ where $T$ is the query
  $\query{q(\tilde{Y})}{c\land\tilde{X}=\tilde{a}}$.

  As we assumed~(\ref{eq1-theo-DN-iff-DNlog}),
  we have to establish that $\mathcal{D}_{\mathcal{C}} \models_v
  \forall_{\restrict{\tilde{X}}{\tau}} \big[
  \sat{\restrict{\tilde{X}}{\tau}}{\delta(p)} \rightarrow
  \exists_{\mathcal{Y}}[\sat{\restrict{\tilde{Y}}{\tau}}{\delta(q)}\land c]
  \big]$ holds. Assume $v_1$ is a valuation such that
  \begin{equation}\label{eq2-theo-DN-iff-DNlog}
    \mathcal{D}_{\mathcal{C}} \models_{v_1}
    \sat{\restrict{\tilde{X}}{\tau}}{\delta(p)}
  \end{equation}
  and for each variable $X\not\in\restrict{\tilde{X}}{\tau}$, $v(X)=v_1(X)$.
  By property of $\mathcal{C}$, we can consider the query
  $S':=\query{p(\tilde{b})}{\mathit{true}}$ where 
  $\restrict{\tilde{b}}{\overline{\tau}} = \restrict{\tilde{a}}{\overline{\tau}}$ 
  and $\restrict{\tilde{b}}{\tau}$ is a sequence of ground terms
  such that
  $[\restrict{\tilde{b}}{\tau}] = v_1(\restrict{\tilde{X}}{\tau})$.

  It can be noticed that $S'$ is $\Delta$-more general than $S$.
  As $\Delta$ is DN for $r$, there exists a query $T'$ that is
  $\Delta$-more general than $T$ and such that
  $S'\lra_r T'$. Necessarily,
  $T'=\query{q(\tilde{Y}')}{c'\land \tilde{X}'=\tilde{b}}$ where
  $p(\tilde{X}')\leftarrow c\diamond q(\tilde{Y}')$ is a variant of
  $r$ variable disjoint with $S'$.

  As we assumed~(\ref{eq2-theo-DN-iff-DNlog}), we now have to
  establish that $\mathcal{D}_{\mathcal{C}} \models_{v_1}
  \exists_{\mathcal{Y}}[\sat{\restrict{\tilde{Y}}{\tau}}{\delta(q)}\land c]$
  holds. This is done using the fact that $T'$ is $\Delta$-more general
  than $T$ and that
  $\Set{T'}=\Set{\query{q(\tilde{Y})}{c\land \tilde{X}=\tilde{b}}}$.
  \qed
\end{proof}

\begin{example}
  In the constraint domain  $\mathit{Term}$, DN is equivalent 
  to DNlog.
\end{example}


\subsection{A Syntactic Characterization of Derivation Neutral Filters}
In~\cite{Mesnard02a}, we gave,
in the scope of logic programming, a syntactic
definition of neutral arguments.
Now we extend this syntactic criterion to the more
general framework of constraint logic programming.
First, we need rules in flat form:

\begin{definition}[Flat Rule]
  A rule $r:=p(\tilde{X}) \leftarrow c \diamond
  q(\tilde{Y})$ is said
  to be \emph{flat} if $c$ has the form
  $(\tilde{X}=\tilde{s} \land \tilde{Y}=\tilde{t})$
  where $\tilde{s}$ is a sequence of $\arity{p}$ terms
  and $\tilde{t}$ is a sequence of $\arity{q}$ terms 
  such that $\Var(\tilde{s},\tilde{t})
  \subseteq \mathit{local\_var}(r)$.
\end{definition}
Notice that there are some rules
$r:=p(\tilde{X}) \leftarrow c \diamond q(\tilde{Y})$
for which there exists no ``equivalent'' rule in flat
form. More precisely, there exists no rule
$r':=p(\tilde{X}) \leftarrow c' \diamond q(\tilde{Y})$
verifying $\mathcal{D}_{\mathcal{C}}\models
\exists_{\mathit{local\_var}(r)} c
\leftrightarrow \exists_{\mathit{local\_var}(r')} c'$ (take for instance
$r:=p(X) \leftarrow X>0 \diamond p(Y)$ in 
$\mathcal{R}_{\mathit{lin}}$.)

Syntactic derivation  neutrality is defined that way:
\begin{definition}[Syntactic Derivation Neutral]
  \label{def-syn-DN}
  Let $\Delta:=(\tau,\delta)$ be a filter and
  $r:=p(\tilde{X}) \leftarrow (\tilde{X}=\tilde{s} \land
  \tilde{Y}=\tilde{t})
  \diamond q(\tilde{Y})$
  be a flat rule.
  We say that $\Delta$ is \emph{DNsyn} for $r$ if
  \begin{itemize}
  \item \textrm{\bf(DNsyn1)}
    $\restrict{\query{p(\tilde{s})}{\mathit{true}}}{\tau}$ is more
    general than $\delta(p)$,
  \item \textrm{\bf(DNsyn2)}
    $\delta(q)$ is more general than
    $\restrict{\query{q(\tilde{t})}{\mathit{true}}}{\tau}$,
  \item \textrm{\bf(DNsyn3)}
    $\Var(\restrict{\tilde{s}}{\tau})\cap
    \Var(\restrict{\tilde{s}}{\overline{\tau}})=\emptyset$,
  \item \textrm{\bf(DNsyn4)}
    $\Var(\restrict{\tilde{s}}{\tau})\cap
    \Var(\restrict{\tilde{t}}{\overline{\tau}})=\emptyset$.
  \end{itemize}
\end{definition}

\begin{example}
  In Example~\ref{example-DNlog-Term}, the rule $r$ is flat.
  Moreover, the filter $\Delta$ is DNsyn for $r$.
\end{example}

A connection between DN, DNsyn and DNlog is as follows:
\begin{proposition}\label{prop-DNsyn-implies-DN}
  \label{prop-DNsyn-DN}
  Let $r$ be a flat rule and $\Delta$ be a filter.
  If $\Delta$ is DNsyn for $r$ then $\Delta$ is DNlog for $r$
  hence (by Proposition~\ref{prop-DNlog-implies-DN})
  $\Delta$ is DN for $r$. If $\Delta$ is DNlog for $r$
  then {\bf(DNsyn1)} holds.
\end{proposition}

Notice that a DNlog filter is not necessarily DNsyn because
one of {\bf(DNsyn2--4)} may not hold:
\begin{example}
  In $\mathcal{R}_{\mathit{lin}}$, consider the
  flat rule $r$:
  \[p(X_1,X_2) \leftarrow X_1=A\land Y_1=A\land X_2=A-A\land
  Y_2=A-A \diamond p(Y_1,Y_2)\;.\]
  Let $\Delta:=(\tau,\delta)$ where $\tau(p)=\{1\}$ and
  $\delta(p)=\query{\restrict{p}{\tau}(X)}{X\geq 0}$.
  Then, $\Delta$ is DNlog for $r$, but none of
   {\bf(DNsyn2--4)} hold.
\end{example}

However, in the special case of logic programming, we have:
\begin{proposition}[Logic Programming]
  \label{theo-DN-DNsyn-log-prog}
  Suppose that $\mathcal{C}=\mathit{Term}$.
  Let $r$ be a flat rule and $\Delta$ be filter.
  If $\Delta$ is DNlog for $r$ then {\bf(DNsyn3)} and
  {\bf(DNsyn4)} hold.
\end{proposition}


\section{Connections with Earlier Results}
\label{section-special-kind-filter}

The results of~\cite{Payet04b} can be easily obtained within
the framework presented above.
It suffices to consider the following special kind of filter:
%
\begin{definition}[Open Filter]
  \label{def-does-not-filter} 
  We say that $\Delta:=(\tau,\delta)$
  is an \emph{open filter} if for all $p\in\Pi'_L$,
  $\delta(p)$ has the form
  $\query{\restrict{p}{\tau}(\tilde{Z})}{\mathit{true}}$
  where $\tilde{Z}$ is a sequence of distinct variables.
\end{definition}

In an open filter, the function
$\delta$ ``does not filter anything":
\begin{lemma}\label{lemma-open-filters1}
  Let $\Delta:=(\tau,\delta)$ be an open filter. Then,
  a query $S'$ is $\Delta$-more general than a
  query $S$ if and only if $\restrict{S'}{\overline{\tau}}$
  is more general than $\restrict{S}{\overline{\tau}}$.
\end{lemma}
Consequently, an open filter is uniquely
determined by its set of positions.
When reconsidering the definitions
and results of the preceding section within such a context,
we exactly get what we presented in~\cite{Payet04b}.
In particular, Definition~\ref{def-log-DN} can be rephrased as:
%
\begin{definition}[Logical Derivation Neutral]
  \label{def-DNlog-open-filter}
  A set of positions $\tau$ is \emph{DNlog} for a rule
  $r:=p(\tilde{X})\leftarrow c\diamond q(\tilde{Y})$ if
  $\mathcal{D}_{\mathcal{C}} \models 
  c\rightarrow \forall_{\restrict{\tilde{X}}{\tau}}
  \exists_{\mathcal{Y}} c$
  where $\mathcal{Y}:=\Var(\restrict{\tilde{Y}}{\tau}) \cup
  \mathit{local\_var}(r)$.
\end{definition}

As stated in Sect~\ref{section-intro},
the framework presented in this paper is a strict generalization
of that of \cite{Payet04b}.
This is illustrated by the following example.
\begin{example}[Example~\ref{example-DNlog-Term} continued]
  \label{ex-stric-generalization}
  First, notice that, as $\query{p(Y)}{c}$ is not more general than
  $\query{p(X)}{c}$, Corollary~\ref{coro-p-if-p}
  does not allow to infer that $\query{p(X)}{c}$ loops
  with respect to $\{r\}$.

  Let us try to use Definition~\ref{def-DNlog-open-filter}
  to prove that the argument of $p$ is ``irrelevant''. We
  let $\tau(p)=\{1\}$. Hence, $\restrict{\tilde{X}}{\tau}=X$,
  $\restrict{\tilde{Y}}{\tau}=Y$ and $\mathit{local\_var}(r)=\{A\}$.
  Let us consider a valuation $v$ such that $v(X)=f(a)$, $v(Y)=f(f(a))$
  and $v(A)=a$. So, we have $\mathcal{D}_{\mathcal{C}} \models_v c$.
  But we do not have $\mathcal{D}_{\mathcal{C}} \models_v
  \forall_{\restrict{\tilde{X}}{\tau}} \exists_{\mathcal{Y}} c$.
  For instance, if we consider $v_1$ such that $v_1(X)=a$ and
  $v_1(Z)=v(Z)$ for each variable $Z$ distinct from $X$, we do
  not have $\mathcal{D}_{\mathcal{C}} \models_{v_1}
  \exists_{\mathcal{Y}} c$ as the subformula $X=f(A)$ of $c$
  cannot hold, whatever value is assign to $A$.
  Consequently, we do not have $\mathcal{D}_{\mathcal{C}} \models_v
  c \rightarrow \forall_{\restrict{\tilde{X}}{\tau}} \exists_{\mathcal{Y}} c$,
  so $\tau$ is not DNlog for $r$. As $\mathcal{C}=\mathit{Term}$,
  by Theorem~\ref{prop-DN-implies-DNlog2} $\tau$ is not DN for $r$.
  Therefore, using open filters with Proposition~\ref{propo-p-if-p-Delta}
  we are not able to prove that
  $\query{p(X)}{c}$ loops with respect to $\{r\}$.

  However, in Example~\ref{example-DNlog-Term}, we noticed that any
  filter $\Delta:=(\tau,\delta)$ where $\tau(p)=\{1\}$
  and $\delta(p)=\query{\restrict{p}{\tau}(X)}{X=f(A)}$ is DNlog,
  hence DN, for $r$. Moreover, for such a filter, $\query{p(Y)}{c}$ is
  $\Delta$-more general than $\query{p(X)}{c}$. Consequently,
  by Proposition~\ref{propo-p-if-p-Delta},
  $\query{p(X)}{c}$ loops with respect to $\{r\}$.
\end{example}
\section{Conclusion}

We have presented a criterion to detect non-terminating atomic 
queries with respect to a binary CLP clause. This criterion generalizes 
our previous papers in the CLP settings and allows us to reconstruct the work
we did in the LP framework. However, when switching from LP to CLP,
we lose the ability to compute, given a binary clause, a useful filter.
We plan to work on this and try to define some conditions on the
constraint domain which enable the computation of such filters.
Moreover, as pointed out by an anonymous referee, DNsyn and DNlog seem to
be independent notions which we proved to coincide only for open filters
with the specific constraint domain $\mathit{Term}$.
In Theorem~\ref{prop-DN-implies-DNlog2} we investigate the relationship
between DNlog and DN while Proposition~\ref{prop-DNsyn-implies-DN} and
Proposition~\ref{theo-DN-DNsyn-log-prog} essentially establish some
connections between DNsyn and DNlog. The study of relationship between
DNsyn and DN is still missing and we intend to work on this shortly.



\end{document}